\documentclass[fleqn,usenatbib]{mnras}

\usepackage{amsmath}
\usepackage{tabularx}

\usepackage[T1]{fontenc}

\DeclareRobustCommand{\VAN}[3]{#2}
\let\VANthebibliography\thebibliography
\def\thebibliography{\DeclareRobustCommand{\VAN}[3]{##3}\VANthebibliography}

\usepackage{graphicx}	
\usepackage{amsmath}	
\usepackage{amssymb}	

\newcommand{\hi } {{\rm H}\,{\small\rm I}}
\newcommand{\hii } {{\rm H}\,{\small\rm II}}
\newcommand{\nii } {[{\rm N}\,{\small\rm II}]}

\begin{document}
\title[The star-formation relation in TDGs]{Molecular and Ionized Gas in Tidal Dwarf Galaxies: The Spatially Resolved Star-Formation Relation}

 \author[N. Kovakkuni et al.]{N. Kovakkuni$^{1}$\thanks{E-mail: navyasree.kovakkuni@ua.cl}, F. Lelli$^{2}$\thanks{E-mail: federico.lelli@inaf.it}, P.-A. Duc$^{3}$, M. Boquien$^{1}$, J. Braine$^{4}$, E. Brinks$^{5}$, V. Charmandaris$^{6, 7, 8}$, \newauthor F. Combes$^{9}$, J. Fensch$^{10}$, U. Lisenfeld$^{11,12}$, S. S. McGaugh$^{13}$, J. C. Mihos$^{13}$, M. S. Pawlowski$^{14}$, \newauthor Y. Revaz$^{15}$, P. M. Weilbacher$^{14}$
 \\
 $^{1}$Centro de Astronom\'ia (CITEVA), Universidad de Antofagasta, Avenida Angamos 601, Antofagasta, Chile\\
$^{2}$INAF, Arcetri Astrophysical Observatory, Largo Enrico Fermi 5, 50125, Florence, Italy\\
 $^{3}$Universit\'e de Strasbourg, CNRS, Observatoire astronomique de Strasbourg (ObAS), UMR 7550, 67000 Strasbourg, France\\
 $^{4}$Laboratoire d’Astrophysique de Bordeaux, Univ. Bordeaux, CNRS, B18N, all\'ee Geoffroy Saint-Hilaire, 33615 Pessac, France\\
$^{5}$Centre for Astrophysics Research, University of Hertfordshire, College Lane, Hatfield AL10 9AB, UK\\
 $^{6}$Department of Physics, University of Crete, Heraklion, 71003, Greece\\
$^{7}$Institute of Astrophysics, Foundation for Research and Technology-Hellas (FORTH), Heraklion, 70013, Greece\\
$^{8}$School of Sciences, European University Cyprus, Diogenes street, Engomi, 1516 Nicosia, Cyprus\\
$^{9}$LERMA, Observatoire de Paris, PSL Research Universit\'e, CNRS, Sorbonne Universit\'e, UPMC, Paris, France\\
$^{10}$Univ. Lyon, ENS de Lyon, Univ. Lyon 1, CNRS, Centre de Recherche Astrophysique de Lyon, UMR5574, 69007 Lyon, France\\
$^{11}$Departamento de F\'isica Te\'orica y del Cosmos, Universidad de Granada, 18071 Granada, Spain \\
$^{12}$ Instituto Carlos I de F\'isica Te\'orica y Computacional, Facultad de Ciencias, 18071 Granada, Spain \\
 $^{13}$Department of Astronomy, Case Western Reserve University, Cleveland OH 44106, USA\\
 $^{14}$Leibniz-Institut f\"ur Astrophysik Potsdam (AIP), An der Sternwarte 16, 14482 Potsdam, Germany\\
 $^{15}$Institute of Physics, Laboratory of Astrophysics, \'Ecole Polytechnique F\'ed\'erale de Lausanne (EPFL), 1290 Sauverny, Switzerland\\
 }




\label{firstpage}
\pagerange{\pageref{firstpage}--\pageref{lastpage}}

\maketitle


\begin{abstract}    

Tidal dwarf galaxies (TDGs) are low-mass objects that form within tidal and/or collisional debris ejected from more massive interacting galaxies. We use CO($1-0$) observations from ALMA and integral-field spectroscopy from MUSE to study molecular and ionized gas in three TDGs: two around the collisional galaxy NGC\,5291 and one in the late-stage merger NGC\,7252. The CO and H$\alpha$ emission is more compact than the \hi\ emission and displaced from the \hi\ dynamical center, so these gas phases cannot be used to study the internal dynamics of TDGs. We use CO, \hi, and H$\alpha$ data to measure the surface densities of molecular gas ($\Sigma_{\rm mol}$), atomic gas ($\Sigma_{\rm atom}$) and star-formation rate ($\Sigma_{\rm SFR}$), respectively. We confirm that TDGs follow the same spatially integrated $\Sigma_{\rm SFR}-\Sigma_{\rm gas}$ relation of regular galaxies, where $\Sigma_{\rm gas} = \Sigma_{\rm mol} + \Sigma_{\rm atom}$, even though they are \hi\ dominated. We find a more complex behaviour in terms of the spatially resolved $\Sigma_{\rm SFR}-\Sigma_{\rm mol}$ relation on sub-kpc scales. The majority ($\sim$60$\%$) of SF regions in TDGs lie on the same $\Sigma_{\rm SFR}-\Sigma_{\rm mol}$ relation of normal spiral galaxies but show a higher dispersion around the mean. The remaining fraction of SF regions ($\sim$40$\%$) lie in the starburst region and are associated with the formation of massive super star clusters, as shown by Hubble Space Telescope images. We conclude that the local SF activity in TDGs proceeds in a hybrid fashion, with some regions comparable to normal spiral galaxies and others to extreme starbursts.
\end{abstract}




\begin{table*}
\centering
\caption{TDG sample. Distances are adopted from \citet{lelli2015gas}. Redshifts and systemic velocities are computed from H$\alpha$ emission lines (Sect.\,\ref{sec:lines}). SFRs, molecular gas masses, and atomic gas masses are computed within the total CO emitting area, but the TDG size is significantly larger (see Sect.\,\ref{sec:gasmass} for details).}\label{tab:gp}
\begin{tabular}{cccccccccc}
 \hline
Galaxy & R.A.    & Dec.   & Dist.  &  $z$ & V$_{\rm sys}$  & SFR  & M$_{\rm mol}$ & M$_{\rm atom}$ &  Area\\
       & (J2000) & (J2000)& (Mpc) &  &   (km s$^{-1}$) & (M$_{\odot}$  yr$^{-1}$) & (10$^7$ M$_{\odot}$) & (10$^7$ M$_{\odot}$) & (kpc$^2$)\\
\hline 
NGC 5291N & 13 47 20.3 & -30 20 54 & 62 &  0.014 & 4228.8  & 0.6$\pm$0.1 & 5.5$\pm$1.1 & 163.6$\pm$16.7 & 25.0\\ 
NGC 5291S & 13 47 22.7 & -30 27 40  & 62 &  0.016 & 4780.4  & 0.2$\pm$0.1 & 2.6$\pm$0.4 & 124.1$\pm$16.1 & 16.0\\
NGC 7252NW  & 22 20 33.7 & -24 37 24 & 66.5 & 0.016 & 4771.6 & 0.06$\pm$0.02 & 3.2$\pm$0.6 & 11.7$\pm$2.2 & 12.3 \\
\hline
\end{tabular}
\end{table*}

\section{Introduction}\label{sec:intro}

The process of star formation (SF) plays a key role in the formation and evolution of galaxies. Key insights into the SF process are given by empirical relations that connect the star formation rate (SFR) of a galaxy to the availability of gas in the interstellar medium (ISM). One such relation is the Kennicutt-Schmidt (KS) relation. In its original form, the \citet{schmidt1959apj} relation connected volume densities of SFR ($\rho_{_{\rm SFR}}$) and atomic gas mass ($\rho_{_{\rm HI}}$) in star-forming regions of the Milky Way:
\begin{equation}\label{eq:Schmidt}
     \rho_{_{\rm SFR}} \propto \rho_{_{\rm HI}}^n.
\end{equation}
Subsequent studies \citep{kennicutt1998global} of star-forming galaxies revealed a tight relation between the disk-averaged SFR surface densities ($\Sigma_{\rm SFR}$) and total (atomic plus molecular) gas mass surface densities ($\Sigma_{\rm gas}$):
\begin{equation}\label{eqn:sfrelation}
   \Sigma_{\rm SFR} = A\, \Sigma_{\rm gas}^N.
\end{equation}

With the advent of multi-wavelength observations at high angular resolution, it has become possible to study the KS relation in a spatially resolved fashion on kpc-scales in a variety of environments, from spiral galaxies to interacting objects \citep[e.g.,][]{bigiel2008star,leroy2008star, Boquien2011}. These works suggested that the SFR surface density correlates more strongly with molecular gas (H$_2$) than atomic gas in the inner H$_2$-dominated regions of star-forming disks. The situation, however, remains unclear in the \hi-dominated regime, typical of dwarf galaxies \citep{roychowdhury2014relation,roychowdhury2015spatially} as well as the outermost parts of spiral galaxies \citep{Bigiel2010}. In this regime, flares in the outer gaseous disks (i.e., a radial increase of the disk thickness) may play an important role \citep{Bacchini2019, Bacchini2020}, suggesting that the volume density of total gas (atomic plus molecular) best correlates with the volume density of SFR, in analogy to the original form in Eq.\,\ref{eq:Schmidt}.

A key question is whether every galaxy follows the same KS relation, thus whether the SF process is ``universal'' or not on sub-kpc scales. For example, there is a clear link between SF activity and galaxy interactions \citep[e.g.,][]{Ellison2013}. Tidal interactions lead to gas inflows towards the galaxy centers, which can temporarily enhance the SFRs, and move the resulting systems above the mean KS relation in the so-called starburst regime \citep[e.g.][]{barnes1991fueling, Renaud2014, Ellison2020}. At the same time, galaxy collisions expel gas and stars into intergalactic space, leading to the formation of tails and bridges in which new stars can form. SF, indeed, can occur within gas debris surrounding interacting systems, sometimes even at 100 kpc away from the parent galaxies \citep{mirabel1992genesis, duc:2006, boquien2007polychromatic, Boquien2011}. Is the SF occurring in such extreme environments proceeding in a similar way as in galaxy disks?

Tidal dwarf galaxies (TDGs) are self-gravitating objects found within tidal debris, which show in-situ SF and have masses and sizes comparable to ``normal'' dwarf galaxies (\citealt{duc:2006}). \citet{zwicky1956multiple} was the first to suggest the possible formation of TDGs around interacting galaxies; this hypothesis was later confirmed by several observational and theoretical studies (e.g. \citealt{mirabel1992genesis,  barnes1992formation, elmegreen1993interaction}). Hereafter, for simplicity, we will use the term TDG to also include newborn galaxies that form within collisional debris (rather than tidal ones), such as the SF complexes in the collisional ring of NGC\,5291 \citep{Bournaud2007}. In addition, we will refer to ``bona-fide TDGs'' to indicate systems that show internal kinematics decoupled from the surrounding debris, pointing to self-gravity within a local potential well \citep[e.g.,][]{lelli2015gas}. Bona-fide TDGs have been identified around several interacting systems \citep[e.g.,][]{duc:2006, lelli2015gas}. 

TDGs form out of gas that has been pre-enriched in their parent massive galaxies, giving them higher gas-phase metallicities (about $0.3-0.5 Z_{\odot}$) than ``classical'' dwarf galaxies \citep{duc1998young}. Thus, contrarily to most dwarf galaxies, TDGs are easily detected in the CO emission line, and it is sensible to use the Milky-Way $X_{\rm CO}$ factor to convert the observed CO flux density into H$_2$ column densities \citep{braine2001abundant, lisenfeld2016molecular, querejeta2021}. \citet{braine2001abundant} used single-dish CO observations of eight TDGs to conclude that they follow the same galaxy-averaged KS relation of usual galaxies. \citet{Boquien2011} reached the same conclusion for a TDG candidate in the interacting system Arp\,158. On the other hand, \citet{lisenfeld2016molecular} presented a spatially resolved study of a TDG in the Virgo Cluster and found that it lies below the mean KS relation. Finally, \citet{querejeta2021} analysed high-resolution ALMA observations of a TDG in the interacting system Arp\,94, and found that it may lie either on or off the KS relation, depending on whether one considers the whole CO flux or only the one associated to giant molecular clouds. The diverse outcome of these works may point to different behaviour in the galaxy-averaged and spatially-resolved KS relation and/or intrinsic differences in the SF properties of individual TDGs, potentially related to their different formation and evolutionary histories. New studies are needed to clarify the situation.

In this paper, we study the spatially resolved KS relation in a sample of three TDGs: NGC\,5291N, NGC\,5291S, and NGC\,7252NW. We probe their molecular gas content using CO(1-0) observations from the Atacama Large Millimeter/submillimeter Array (ALMA), and their ionized gas content (H$\alpha$ and H$\beta$ emission) using integral-field spectroscopy (IFS) from the Multi-Unit Spectroscopic Explorer (MUSE) mounted on the Very Large Telescope (VLT). New observations and ancillary data are described in Sect.\,\ref{sec:obs}. Results on the gas distribution and kinematics as well as on the KS relation are presented in Sect.\,\ref{sec:res}. Finally, we summarize our findings in Sect.\,\ref{sec:conc}

\section{Observations}\label{sec:obs}

\subsection{Galaxy sample}

We obtained high-resolution ALMA observations for three TDGs that were selected from the sample of \citet{lelli2015gas}, based on the availability of single-dish CO fluxes and high-quality \hi\ maps. Two of them are part of the NGC\,5291 system; the remaining one is part of the NGC\,7252 merger. The location of these TDGs within the overall structure of the parent system can be appreciated in Figure\,1 of \citet{lelli2015gas}. Table\,\ref{tab:gp} lists the general properties of these TDGs.

NGC\,5291 is a perturbed early-type galaxy that is surrounded by a giant \hi\ ring \citep{malphrus1997ngc}, suggesting a past head-on collision \citep{Bournaud2007}. The system is located in the outer region of the galaxy cluster Abell 3574. Early studies by \citet{longmore1979ngc} in the collisional ring of NGC 5291 revealed the presence of star-forming regions out to 100 kpc from the central galaxy. Subsequently, \citet{duc1998young} found that these star-forming complexes have similar sizes and SFRs of dwarf galaxies but higher metallicities, suggesting a TDG origin. 
Three star-forming complexes (NGC\,5291N, NGC\,5291S and NGC\,5291SW) are associated with strong \hi\ concentrations that display a velocity gradient decoupled from the underlying collisional material, pointing to rotation within a self-gravitating potential well \citep{Bournaud2007, lelli2015gas}. In this paper, we focus on NGC\,5291N and NGC\,5291S because no single-dish CO observations are available for NGC\,5291SW.

NGC\,7252 (also known as ``Atoms for Peace'') is a late-stage merger remnant with two gas-rich tidal tails extending to the east and north-west. Numerical simulations \citep{borne1991merger,hibbard1995dynamical, chien2010dynamically} were able to reproduce the observed morphology and kinematics of NGC\,7252 from the merging of two disk galaxies. \citet{hibbard1994cold} identified two TDG candidates in the North-Western and Eastern tails (NGC\,7252NW and NGC\,7252E). These star-forming complexes were confirmed as bona-fide TDGs based on their H$\alpha$ kinematics \citep{Bournaud2004}, \hi\ kinematics \citep{lelli2015gas} and relatively high gas metallicities \citep{lelli2015gas}. In this paper, we focus on NGC\,7252NW because no single-dish CO observations are available for NGC\,7252SE.

\subsection{ALMA data}

The three TDGs were observed by the ALMA 12m array in January 2016 (Project 2015.1.00645.S; PI: F. Lelli). The time on source was about 2.2 hrs for both NGC\,5291N and NGC\,5291S, and about 4.5 hrs for NGC\,7252NW. We used ALMA band 3 with a mixed spectral setup, using four spectral windows with a bandwidth of 1875 MHz each. A high-resolution spectral window was centered at the frequency of the redshifted CO($1-0$) line and covered with 3480 channels, providing a spectral resolution of 976.6 kHz ($\sim$2.6 km s$^{-1}$). Three low-resolution spectral windows were centered around 99, 100, and 110 GHz to target the mm continuum; they were covered with 128 channels providing a spectral resolution of 31.250 MHz (ranging from $\sim$84 to $\sim$94 km s$^{-1}$). The observations were pointed at the \hi\ kinematic center of the TDGs and have a field of view of $\sim$50$''$, set by the full-width half-maximum (FWHM) of the primary beam.

\begin{table}
\centering
\caption{Properties of CO data cubes with a channel width of $\sim$5 km\,s$^{-1}$.}\label{tab:cubes}
\begin{tabular}{cccccccc}
 \hline
Galaxy & Beam   &  Beam PA & $\sigma_{\rm cube}$ \\
       & (arcsec$\times$arcsec) & (degrees) & (mJy beam$^{-1}$)\\
\hline 
NGC 5291N & $2.1\times1.7$ & 75.5 & 0.5 \\ 
NGC 5291S & $2.1\times1.8$ & -4.6 & 0.5 \\
NGC 7252NW  & $2.7\times1.6$ & -8.3 & 0.5 \\
\hline
\end{tabular}
\end{table}
The data reduction was performed with the Common Astronomy Software Applications (\textsc{Casa}) package \citep{CASA}. The $uv$ data were flagged and calibrated using the standard \textsc{Casa} pipeline. Both continuum and line data were imaged using the \texttt{tclean} task with a H\"ogbom deconvolver and Briggs weighting with a robust parameter of 0.5. Continuum images were constructed  by combining all four spectral windows, excluding channels with line emission. Continuum emission was detected only in NGC\,5291NW and subtracted from the CO($1-0$) line channels using the task \texttt{uvcontsub}. The properties of the CO($1-0$) line cubes are summarized in Table\,\ref{tab:cubes}. In particular, the spatial resolution (FWHM of the synthesized beam) is about 2$''$ which corresponds to about 600 pc and 650 pc at the distances of NGC\,5291 and NGC\,7252, respectively.

\begin{figure*}
    \centering
    \includegraphics[width=0.96\textwidth]{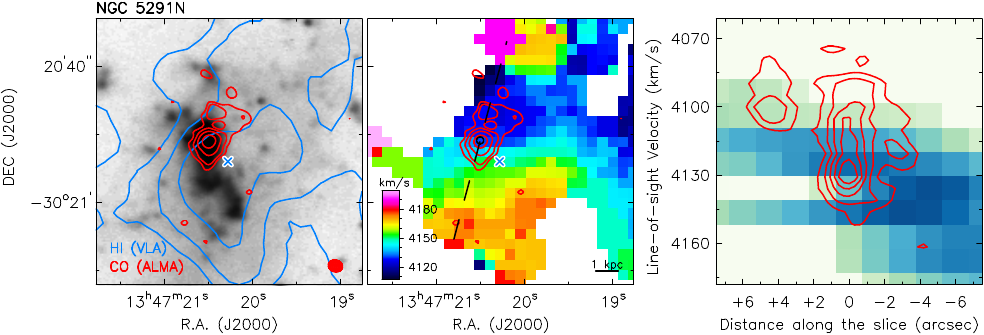}
    \includegraphics[width=0.96\textwidth]{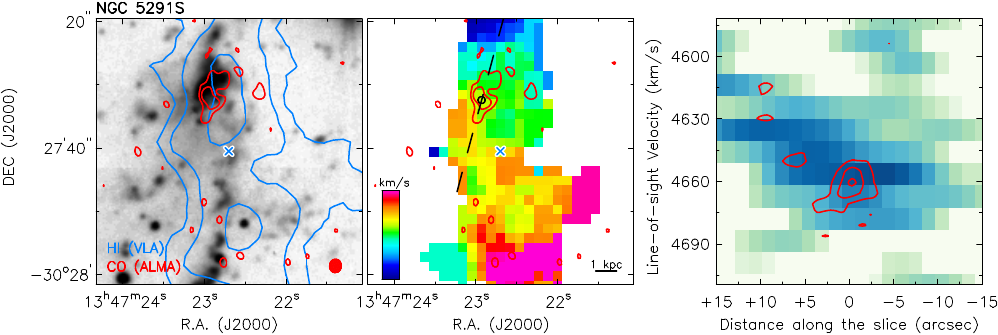}
    \includegraphics[width=0.96\textwidth]{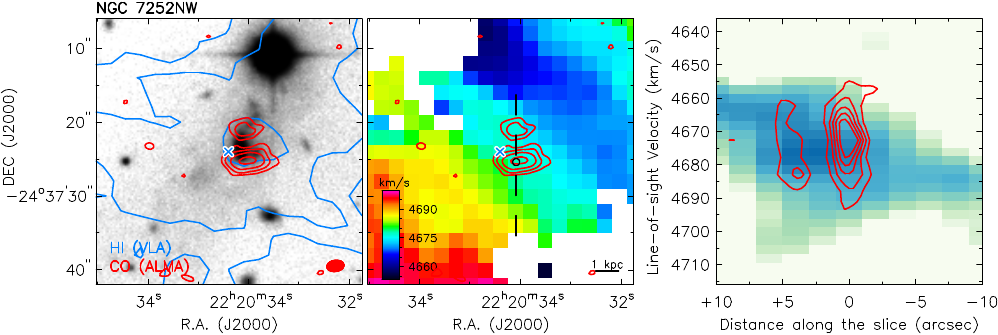}
    \caption{Gas distribution and kinematics in NGC 5291N (top), NGC 5291S (middle), and NGC 7252NW (bottom). In all panels, the CO($1-0$) data are from this work, while the other data come from \citet{lelli2015gas}. \textit{Left panels:} optical R$-$band image overlaid with the \hi\ map (blue contours) and the CO($1-0$) map (red contours). \hi\ contours are the same as in \citet{lelli2015gas}. CO contours are at (3, 6, 12, 24) $\sigma_{\rm map}$, where $\sigma_{\rm map}=0.02$ Jy beam$^{-1}$ km s$^{-1}$ for NGC\,7252NW and $\sigma_{\rm map}=0.03$ Jy beam$^{-1}$ km s$^{-1}$ for both NGC\,5291N and NGC\,5291S. The cross shows the \hi\ kinematic center which was chosen as the ALMA pointing center. The ALMA beam is shown by the red ellipse to the bottom-right corner; the \hi\ beam (not shown) is about 4 times larger for NGC\,5291 and 6 times for NGC\,7252. \textit{Middle panels:} \hi\ velocity field overlaid with the CO map. The dashed line shows the slit used to extract the PV diagram. The physical scale of 1 kpc is indicated by the bar in the bottom-right corner. \textit{Right panels}: PV diagrams from the \hi\ cube (blue colorscale) overlaid with those from the CO cube (red contours). CO contours range from 3$\sigma_{\rm cube}$ to 15$\sigma_{\rm cube}$ in steps of 3$\sigma_{\rm cube}$ (see Table\,\ref{tab:cubes}).}
    \label{fig:overview}
\end{figure*}
CO intensity (moment-zero) maps were constructed by summing channels with CO emission and are shown in Fig.\,\ref{fig:overview}. They are discussed in Sect.\,\ref{sec:CO} for illustrative purpose only: CO fluxes are measured by extracting integrated spectra in various spatial regions as described in Sect.\,\ref{sec:lines}. No corrections for the primary beam are applied in the moment maps, instead they are applied to the derived fluxes as described in Sect.\,\ref{sec:lines}. Since the ALMA primary beam is significantly larger than the CO emitting area, the primary beam attenuation has little effect on the moment-zero maps, with the exception of NGC 5291S. For this TDG, the pointing center was chosen in-between two main SF complexes so that the CO emission in the Northern  complex (the Southern is not detected) is $\sim$10$''$ offset from the pointing center 
(see Fig.\,\ref{fig:overview}). It is also possible that the two SF complexes are two distinct objects, whose individual kinematics cannot be discerned with the available \hi\ observations \citep{lelli2015gas}. Correction for primary beam attenuation is described in Sect.\,\ref{sec:lines}.

The CO emission is very compact and confined to one or two major clumps that display no appreciable velocity gradients (Fig.\,\ref{fig:overview}), so moment-one and moment-two maps are not very useful. We will discuss the CO kinematics using position-velocity (PV) diagrams that provide the most direct representation of the 3D data (Sect.\,\ref{sec:CO}). Before extracting integrated spectra, to enhance the signal-to-noise (S/N) ratio of the CO line, we performed Hanning smoothing over three spectral channels, giving a final velocity resolution of $\sim$10 km/s.

To investigate whether the ALMA interferometric observations may be missing diffuse flux on large scales, we compared spatially-integrated CO spectra from ALMA with those from previous single-dish observations \citep{braine2001abundant}. The low S/N of the previous single-dish observations do not allow us to quantify the amount of diffuse molecular gas missed by the ALMA interferometric observations. The comparison, therefore, was inconclusive, but we note that a substantial amount of diffuse CO emission has been found in another TDG (J1023+1952) using ALMA total power (TP) and Atacama Compact Array (ACA) observations in addition to the 12m array \citep{querejeta2021alma}. It is possible, therefore, that our high-resolution observations are probing only the densest CO emission and may be missing some flux on larger scales. ACA and TP observations (or deep IRAM-30 observations) are needed to check this possibility.

\subsection{MUSE data}

NGC\,5291N was observed on 26 June 2014 (Program ID: 60.A-9320; PI: P.-A. Duc) without adaptive optics (AO) for a total exposure time of 1800 sec. These observations are presented in \citet{Fensch2016}, who provides an in-depth study of line ratios and gas metallicity. NGC\,5291N was re-observed on 19 June 2017 during a MUSE AO commissioning run (Program ID: 60.A-9100(G)), but the AO system could not significantly improve on the external seeing, so we will not use these data.

NGC\,5291S was observed on 22 January 2018 as part of Program ID 097.B-0152 (PI: M. Boquien) without AO for a total exposure time of 1800 sec. NGC\,7252NW was observed on 16 July 2017 during a commissioning run of the MUSE AO wide-field mode (Program ID: 60.A-9100(H)) for a total exposure time of 900 sec. All data were reduced using the MUSE pipeline v2.4 and following standard procedures \citep{weilbacher2020}. For NGC\,5291S, the sky was estimated using an offset field, while for NGC\,7252NW, it was estimated within the science exposure. We refer to \citet{Fensch2016} for further details on the data reduction.

\begin{figure*}
    \centering
    \includegraphics[width=\textwidth]{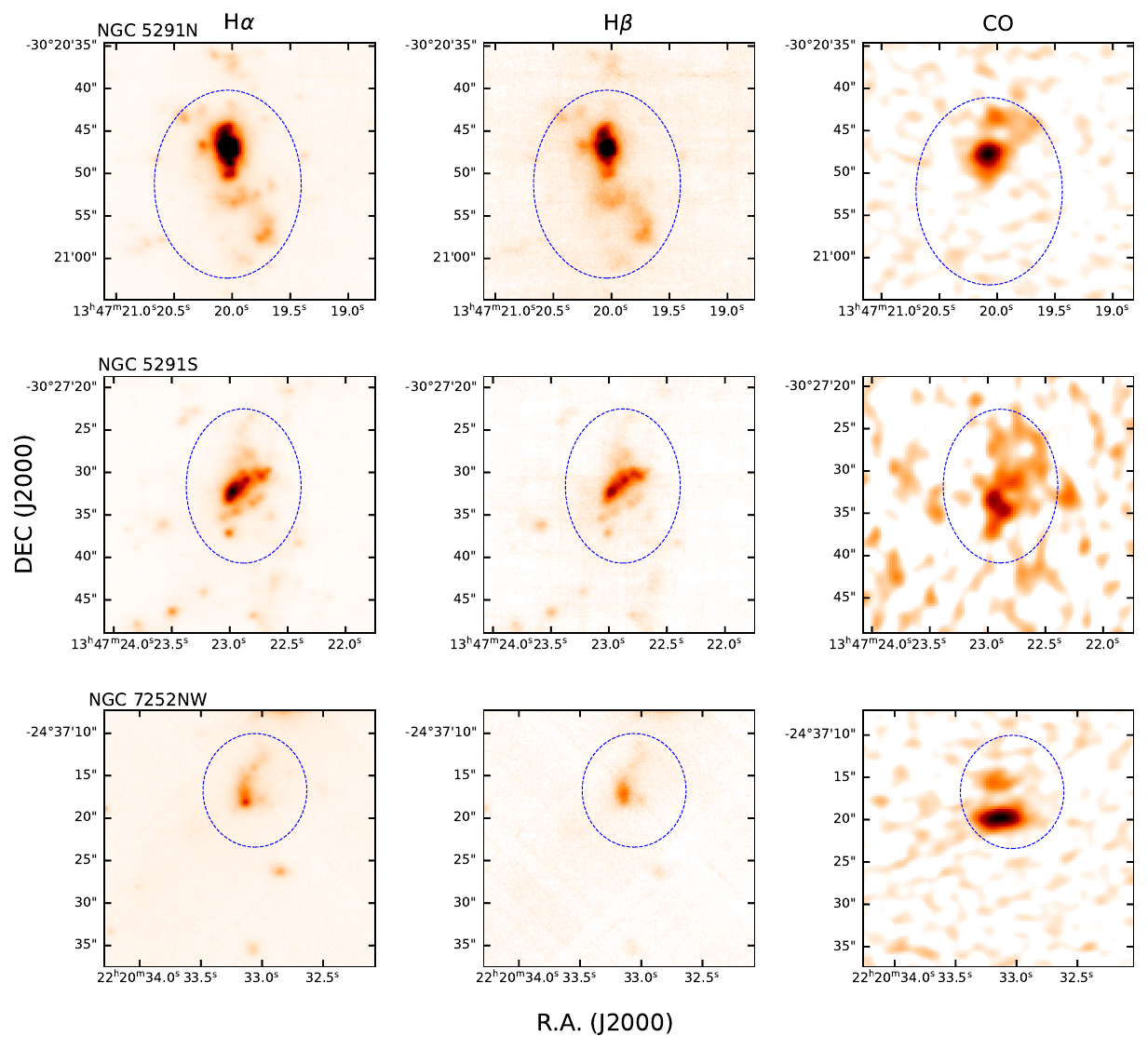}
    \caption{Spatial distribution of H$\alpha$ (left), H$\beta$ (middle), and CO($1-0$) velocity-integrated emission (right) in NGC\,5291N (top), NGC\,5291S (middle) and NGC 7252NW (bottom). The elliptical area marked with blue dashed line shows the integration region used to compute the total H$\alpha$, H$\beta$, and CO fluxes of the TDGs.}
    \label{fig:tot_intensity}
\end{figure*}

Moment-zero maps were constructed by summing channels with H$\alpha$ and H$\beta$ emission. These maps are intended to show the morphology of the ionized gas; line fluxes will be measured by extracting integrated spectra and subtracting the stellar continuum (Sect.\,\ref{sec:lines}).

\begin{figure*}
    \centering
    \includegraphics[width=\textwidth]{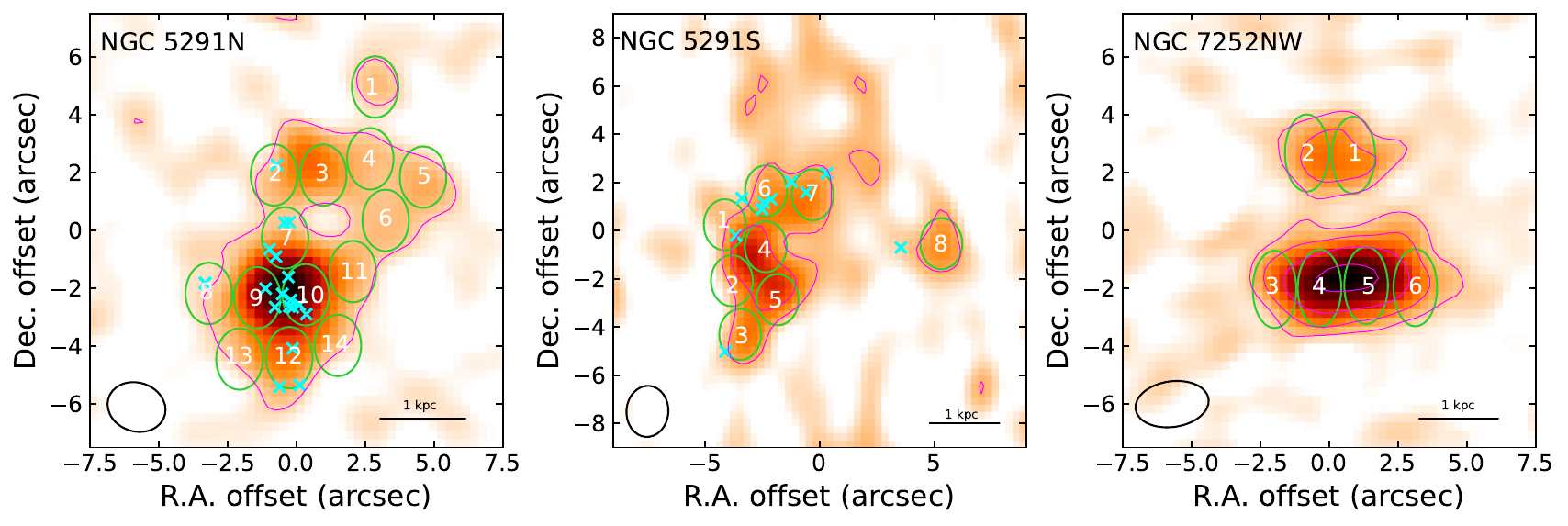}
    \caption{Integration regions for spatially resolved flux measurements in NGC\,5291N (left), NGC\,5291S (middle), and NGC\,7252NW (right). The CO($1-0$) intensity maps (red colorscale) are overlaid with the areas (green ellipses) within which integrated spectra are extracted. In NGC\,5291N and NGC\,5191S, the cyan crosses indicate the location of the ``secure'' massive star clusters identified by \citet{fensch2019massive} using HST images. The size of the ellipse shown in the bottom-left corner is equivalent to the CO($1-0$) beam. Red contours are the same as in Figure\,\ref{fig:overview}.}
    \label{fig:datapoints}
\end{figure*}

\begin{figure*}
    \centering
    \includegraphics[width=\textwidth]{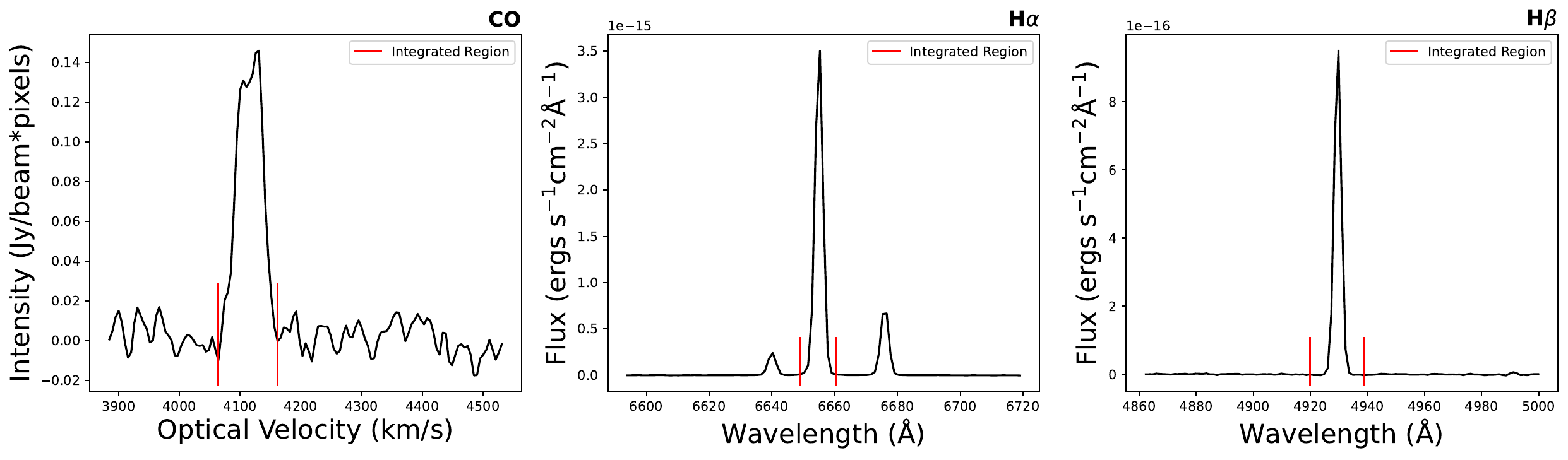}
    \caption{The CO, H$\alpha$+\nii, and H$\beta$ spectra of the star-forming region 9 in NGC\,5291N. Red vertical lines show the range over which the spectra are integrated.}
    \label{fig:e9}
\end{figure*}

\section{Results}\label{sec:res}

\subsection{Distribution and kinematics of multiphase gas}\label{sec:CO}

Figure\,\ref{fig:overview} compares the distribution and kinematics of molecular (CO) and atomic (\hi) gas in our sampled TDGs. In all galaxies, the detected CO emission is much more compact than the \hi\ emission (left panels) and associated with intense SF, as we describe later on. The CO extent is of the order of $1-2$ kpc, while the \hi\ disks have diameters ranging from $\sim$10 to $\sim$15 kpc \citep{lelli2015gas}. The CO emission lies near the edges of the \hi\ peaks, so there is no direct correspondence between the atomic and molecular gas distribution on kpc scales. Moreover, the CO emission is not at the dynamical center of the TDGs, so it cannot be used to trace the underlying large-scale gas rotation (middle panels). The same behaviour is often seen in typical dwarfs and in the outskirts of spirals, where the \hi\ emission traces the object more globally, and there may be regions where atomic gas can cool, condense into H$_2$, and SF can commence \citep[e.g.,][]{2023arXiv230503443H}.

The atomic gas mass of TDGs outweighs the molecular gas mass by factors of $3-30$ within the CO emitting area (see Table\,\ref{tab:gp}) and the stellar mass by factors of $5-15$ \citep[see Table 5 in][]{lelli2015gas}, so the \hi\ center corresponds to the center of mass (in the case of negligible dark matter). Interestingly, position-velocity (PV) diagrams along the CO distribution (right panels) show that there is good agreement between CO and \hi\ line-of-sight velocities, suggesting that the two gas phases are dynamically coupled. The most likely interpretation is that molecular gas is currently forming out of atomic gas, keeping the same kinematics. 

Figure\,\ref{fig:tot_intensity} compares the distribution of molecular and ionized gas, specifically the H$\alpha$ and H$\beta$ lines that will be used to measure the SFRs. Molecular and ionized gas are roughly co-spatial on kpc scales but display different morphologies, so the spatial relation between current star-formation activity and molecular gas reservoir is complex. Importantly, both molecular and ionized gas are much more compact than the atomic gas and similarly displaced from the \hi\ dynamical center. Thus, no firm statement on the large-scale dynamics of TDGs can be inferred from H$\alpha$ emission. The same conclusion was drawn by \citet{lelli2015gas} for NGC 7252NW using H$\alpha$ data from GIRAFFE, which is in good agreement with the new MUSE data. On the contrary, \citet{Flores2016} inferred strong conclusions on the dynamics of the TDGs in NGC\,5291 using H$\alpha$ data from GIRAFFE. Fig.\,\ref{fig:overview} and Fig.\ref{fig:tot_intensity} show that the bright H$\alpha$ emission cannot be used to trace the large-scale gas kinematics of TDGs.

The \hi\ emission appears to be the most extended and most promising tracer to probe the internal dynamics in TDGs, provided that the kinematically decoupled part can be discerned from the tidal tails and properly resolved. The \hi-to-H$_2$ conversion and the most intense SF activity, however, do not occur exactly at the dynamical center, possibly due to variable local conditions such as gas pressure, temperature, and volume density. A similar situation occurs in ``regular'' starburst dwarfs, such as blue compact dwarfs (BCDs), which often show offsets between the peak SF activity and the dynamical center of the galaxy \citep[e.g.,][]{lelli2014}.

\subsection{Emission-Line Measurements}\label{sec:lines}

We combine $\Sigma_{\rm SFR}$ traced by H$\alpha$ emission with $\Sigma_{\rm mol}$ traced by CO emission to investigate the spatially resolved KS relation. We define a set of independent elliptical apertures with major and minor axes matching the FWHM of the ALMA synthesized beam (see Figure \ref{fig:datapoints}). We then extract integrated CO, H$\alpha$, and H$\beta$ spectra within each elliptical aperture from the ALMA and MUSE cubes, respectively. This is nearly equivalent to smoothing and/or re-binning the MUSE data to the lower spatial resolution of the ALMA data. The key advantage of this procedure is to ensure independent measurements of $\Sigma_{\rm SFR}$ and $\Sigma_{\rm mol}$ because the elliptical regions are equal to or larger than the angular resolution. The ellipses are chosen to cover the CO emission down to a contour with $\rm{S/N=3}$; for simplicity, they are oriented in the North-South direction rather than along the PA of the ALMA beam, but this choice has no appreciable effects on the final results. Using a circular aperture would also make no difference in our general results. We have 14, 8, and 6 apertures for NGC\,5291N, NGC\,5291S, and NGC\,7252NW, respectively, for a total of 28 independent measurements. 

We extract the optical and CO($1-0$) spectra of each region using \textsc{Casa}. For each H$\alpha$ and H$\beta$ spectra, we subtract continuum emission by calculating the mean continuum flux from two narrow spectral regions on either side of the emission profile. In all selected regions, the CO($1-0$), H$\alpha$, and H$\beta$ lines are detected with a peak S/N ratio higher than 3.5, apart from region 8 of NGC\,5291S in which H$\beta$ emission is undetected. Figure\,\ref{fig:e9} shows an example of the spectra from one region in NGC\,5291N.

The H$\alpha$ and H$\beta$ lines have a nearly Gaussian shape, but the CO lines do not (see Fig.\,\ref{fig:e9}). Most likely, the Gaussian shape of the MUSE profiles is driven by the instrumental spectral resolution (FWHM of $\sim$80 km s$^{-1}$ around the H$\alpha$ line), while the non-Gaussian shape of CO profiles is intrinsic, given the ALMA spectral resolution of $\sim$10 km s$^{-1}$ (after Hanning smoothing). Rather than fitting a Gaussian function, therefore, we use direct integration to estimate the integrated flux from the emission lines of interest. The starting and ending frequency of integration were defined by visual inspection; they are typically about 100 km s$^{-1}$ wide. In regions with low S/N, it is not trivial to identify the proper integration range. This is especially the case for the H$\alpha$ line because the \nii\ doublet blends with the noise. In such cases, to avoid contamination from \nii\ lines, we estimate the redshifted wavelength of the \nii\ doublet and consider appropriate frequency ranges excluding the \nii\ emission.

To extract CO spectra and measure CO fluxes, we used cubes that are \emph{not} corrected for the ALMA primary-beam attenuation because these cubes have uniform and well-defined noise structure. In primary-beam-corrected cubes, indeed, the noise varies from pixel to pixel, so it is challenging to define the S/N ratio of the line. To recover the correct CO fluxes, instead, we use the primary-beam map and compute the average primary-beam correction within each region, then multiply the uncorrected CO flux by this value. Table \ref{tab2} summarizes all our measurements.

\begin{table*}
\begin{center}
\caption{Line fluxes, molecular masses, and SFRs within the independent regions identified in Fig.\,\ref{fig:datapoints}.}
   \begin{tabular}{cccccccc}
  \hline
   TDG & Region & S$_{\rm CO}$ $\Delta\nu$ & M$_{\rm mol}$  & F$_{\rm H\alpha}$ & F$_{\rm H\beta}$ & Extinction & SFR\\
   & & (Jy km s$^{-1})$ & (10$^6$ M$_{\odot}$) & (10$^{-16}$ ergs s$^{-1}$ cm$^{-2}$) & (10$^{-16}$ ergs s$^{-1}$ cm$^{-2}$)  & (mag) & ($M_{\odot}$ $ yr^{-1}$) \\ [0.2cm]  \hline 
          & 1 & 0.03 & 1.0 & 1.2 & 0.4 & 0.4 & 0.0006\\
            & 2 & 0.04 & 1.4 & 6.0  & 1.9 & 0.3 & 0.003\\
            & 3 & 0.07 & 2.8 & 4.0  & 1.2 &  0.5 & 0.002\\
            & 4 & 0.04 & 1.4 & 1.1  & 0.3 &  0.6 & 0.0007\\
            & 5 & 0.04 & 1.5 & 0.9  & 0.2 &   0.8 & 0.0007\\
            & 6 & 0.04 & 1.5 & 2.1  & 0.6 & 0.5 & 0.001\\
NGC 5291N   & 7 & 0.07 & 2.7 & 65.4 & 18.4&  0.5 & 0.04\\
            & 8 & 0.03 & 1.1 & 15.7 & 4.7 &  0.4 & 0.008\\
            & 9 & 0.2  & 7.4 & 111.6& 29.8&  0.7 & 0.08\\
            & 10& 0.2  & 6.5 & 161.2& 41.1&  0.8 & 0.1\\
            & 11& 0.04 & 1.5 & 7.1  & 2.1 &  0.4 & 0.004\\
            & 12& 0.08 & 3.3 & 66.5 & 17.2&  0.8 & 0.05\\
            & 13& 0.04 & 1.5 & 11.8 & 3.6 &  0.4 & 0.006\\
            & 14& 0.03 & 1.3 & 14.1 & 4.2 &  0.4 & 0.007\\ [0.05cm] \hline 
          \rule{0pt}{3ex}     &1& 0.03 & 1.2 & 55.0 & 14.5 & 0.7 & 0.04\\ 
            &2& 0.05 & 1.9 & 21.6 & 6.0  & 0.6 & 0.01\\
            &3& 0.04 & 1.5 & 7.6  & 2.3  & 0.3 & 0.004\\ 
NGC 5291S   &4& 0.05 & 2.1 & 24.8 & 7.1  & 0.5 & 0.01\\ 
            &5&  0.05 & 2.0& 4.7  & 1.0  & 1.3 & 0.006\\
            &6&  0.04 & 1.4& 28.2 & 8.3  & 0.5 & 0.02\\
            &7& 0.04 & 1.6 & 22.7 & 6.5  & 0.5 & 0.01\\ 
            &8& 0.03 & 1.3 & 0.3  & -    & -   & 0.0001\\ [0.05cm] \hline     
         \rule{0pt}{3ex}     & 1 & 0.03 & 1.5 & 6.9 & 1.9 & 0.6  & 0.005\\
           & 2 & 0.03 & 1.6 & 1.8 & 0.5 & 0.6  & 0.001\\ 
 NGC 7252NW & 3 &0.05 & 2.1 & 0.4 & 0.08& 1.3 & 0.0005\\
            & 4 & 0.1 & 5.2 & 2.0 & 0.5 & 1.0 & 0.002\\
            & 5 & 0.1 & 5.3 & 2.0 & 0.4 & 1.3 & 0.003\\
            & 6 & 0.05& 2.1 & 0.6 & 0.2 & 0.8 & 0.0006\\
  \hline            
            
  \end{tabular}
\end{center}
\label{tab2}
\end{table*}

\subsection{Molecular gas masses \& star-formation rates}\label{sec:gasmass}

We convert the CO line flux of each region (S$_{\rm CO} \Delta\nu$) to a total molecular gas mass (M$_{\rm mol}$, including Helium and heavier elements) assuming the Milky-Way CO-to-$M_{\rm mol}$ conversion factor $\alpha_{\rm CO}$  = 4.3 M$_{\odot}$ (K km s$^{-1}$ pc$^{2}$)$^{-1}$ or equivalently $X_{\rm CO} = 2 \times 10^{20}$ cm$^{-2}$ (K km s$^{-1}$)$^{-1}$. This corresponds to the following equation \citep{bolatto2017edge}:
\begin{equation}\label{eqn:molecular}
    M_{\rm mol} = 1.05 \times 10^4 \, \frac{S_{\rm CO}\Delta\nu \, D_{\rm L}^2}{(1+z)},
\end{equation}
where $M_{\rm mol}$ is in units of M$_{\odot}$, $S_{\rm CO} \Delta\nu$ is in units of Jy km s$^{-1}$, $D_{\rm L}$ is the luminosity distance in Mpc and $z$ is the redshift. The choice of conversion factor in Eq.\,\ref{eqn:molecular} holds for disk galaxies similar to the Milky Way. As discussed in Sect.\,\ref{sec:intro}, the same conversion factor is expected to hold in TDGs because they retain the metallicity of the parent spiral galaxies. Clearly, this is a simplifying assumption because the value of $X_{\rm CO}$ varies even within ``normal'' disk galaxies and is known to increase with decreasing metallicity \citep{bolatto2013co}. The metallicities of our TDGs range from half solar to solar \citep{duc1998young, lelli2015gas}, so the variation in $X_{\rm CO}$ is expected to be null or small, depending on the adopted model \citep[cf.][]{bolatto2013co}. In the worst case scenario, in some star-forming regions of our TDG sample, M$_{\rm mol}$ may be underestimated by a factor of $\sim$2$-$3. The molecular mass surface density ($\Sigma_{\rm mol}$) is derived by dividing M$_{\rm mol}$ by the ellipse area.

The H$\alpha$ flux provides an instantaneous measure of the SFR as nebular emission is produced around young massive stars with masses greater than 10 M$_\odot$ and lifetimes shorter than $10-20$ Myr \citep{kennicutt1998star}. SFR tracers probing longer timescales have been studied and discussed in \citet{boquien2007polychromatic, boquien2009collisional, Boquien2010}. The primary contributor to systematic errors in H$\alpha$-based SFRs is dust extinction, which can be accounted for by using the Balmer decrement  H$\alpha$/H$\beta$. We follow  \citet{bolatto2017edge} to estimate the nebular extinction A$_{\rm H\alpha}$:
\begin{equation}
   A_{\rm H\alpha} = 5.86 \, \log  \, \frac{F_{\rm H\alpha}}{2.86\, F_{\rm H\beta}},
\end{equation}
 where $F_{\rm H\alpha}$ and $F_{\rm H\beta}$ are the integrated fluxes. Here, the case B recombination value of intrinsic Balmer decrement is considered to be 2.86 as suggested by \citet{storey1995recombination} for \hii\ regions at typical electron temperatures and densities. The extinction corrected SFR was computed as
\begin{equation}\label{eqn:sfr}
    {\rm SFR} = 7.9 \times 10^{-42} \, L_{\rm H{\alpha}} \, 10^{({A_{\rm H\alpha}/2.5})},
 \end{equation}
where SFR is in units of M$_{\odot}$ yr$^{-1}$ and $L_{\rm H{\alpha}}$ is the luminosity of H${\alpha}$ in units of ergs s$^{-1}$. This equation assumes solar abundance and a Salpeter initial mass function (IMF) with a mass range of 0.1 to 100 M$_\odot$ \citep{kennicutt1998star}. Finally, we compute $\Sigma_{\rm SFR}$ (in units of M$_{\odot}$ yr$^{-1}$ kpc$^{-2}$) dividing the SFR by the area of the region  (equivalent to the beam size, see Table \ref{tab:cubes}).

\subsection{The Kennicutt-Schmidt relation}\label{sec:ksrelation}

\begin{figure}
    \centering
    \includegraphics[width=0.48\textwidth]{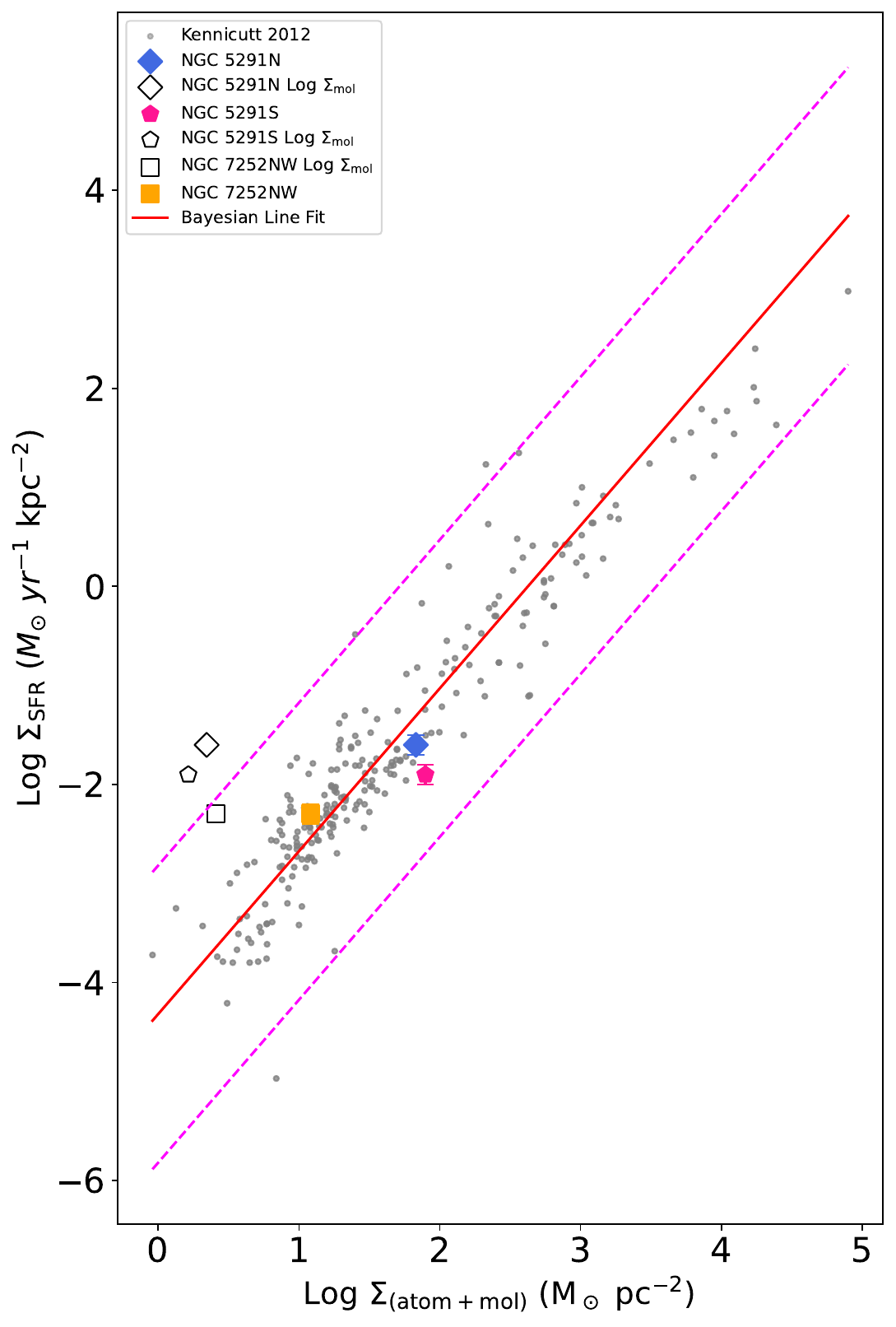}
    \caption{The location of TDGs (blue diamond, yellow square, and pink pentagon) on the spatially integrated Kennicutt-Schmidt relation \citep[grey symbols, from][]{kennicutt2012star}. The open symbols show the location of TDGs if one considers only molecular gas, neglecting atomic gas. The red line shows the best-fit line to the data; the dashed magenta lines correspond to $\pm$3$\sigma_{\rm obs}$ where $\sigma_{\rm obs}$ is the observed vertical scatter.}
    \label{fig:globalKSlaw}
\end{figure}

As a first step, we locate our three TDGs on the spatially integrated KS relation, which compares the total gas surface density (atomic plus molecular) with the SFR. We use the data from \citet{kennicutt2012star} because they are internally self-consistent with our data: the SFRs are measured from extinction-corrected H$\alpha$ fluxes and the molecular gas masses assume the MW $X_{\rm CO}$ factor (cf. with Sect.\,\ref{sec:gasmass}). Figure\,\ref{fig:globalKSlaw} shows that TDGs follow the same KS relation as ``normal'' galaxies, in agreement with earlier results from \citet{braine2001abundant} and \citet{Boquien2011} using single-dish CO observations. Notably, if we consider only molecular gas masses, TDGs would strongly shift to the left of the relation because the total gas mass is heavily dominated by atomic gas, unlike typical spiral galaxies.

Next, we study the spatially resolved SF relation \citep[e.g.,][]{bolatto2017edge,pessa2021star,lin2020almaquest}. We compare our TDGs to 14 spiral galaxies from the ALMA-MaNGA QUEnching and STar formation (ALMaQUEST) survey \citep{lin2019almaquest, lin2020almaquest}. ALMaQUEST data represents the ideal comparison sample because (i) molecular gas masses are estimated using CO($1-0$) data from ALMA as in our work, (ii) SFRs are estimated using extinction-corrected H$\alpha$ fluxes from IFS similar to our work, (iii) the same calibrations have been adopted (Eqs. \ref{eqn:molecular} and \ref{eqn:sfr}), and (iv) the angular resolution of ALMaQUEST data ($\sim$2.5$''$) is similar to that of our data ($\sim$2$''$) albeit the ALMaQUEST physical resolution ranges from 0.5 - 6.5 kpc depending on the galaxy distance while our physical resolution is fixed at $\sim$0.6 kpc. In addition, unlike other samples \citep{bolatto2017edge, pessa2021star}, the ALMaQUEST data covers low gas surface densities similar to those in TDGs.

\begin{figure}
    \centering
    \includegraphics[width=0.48\textwidth]{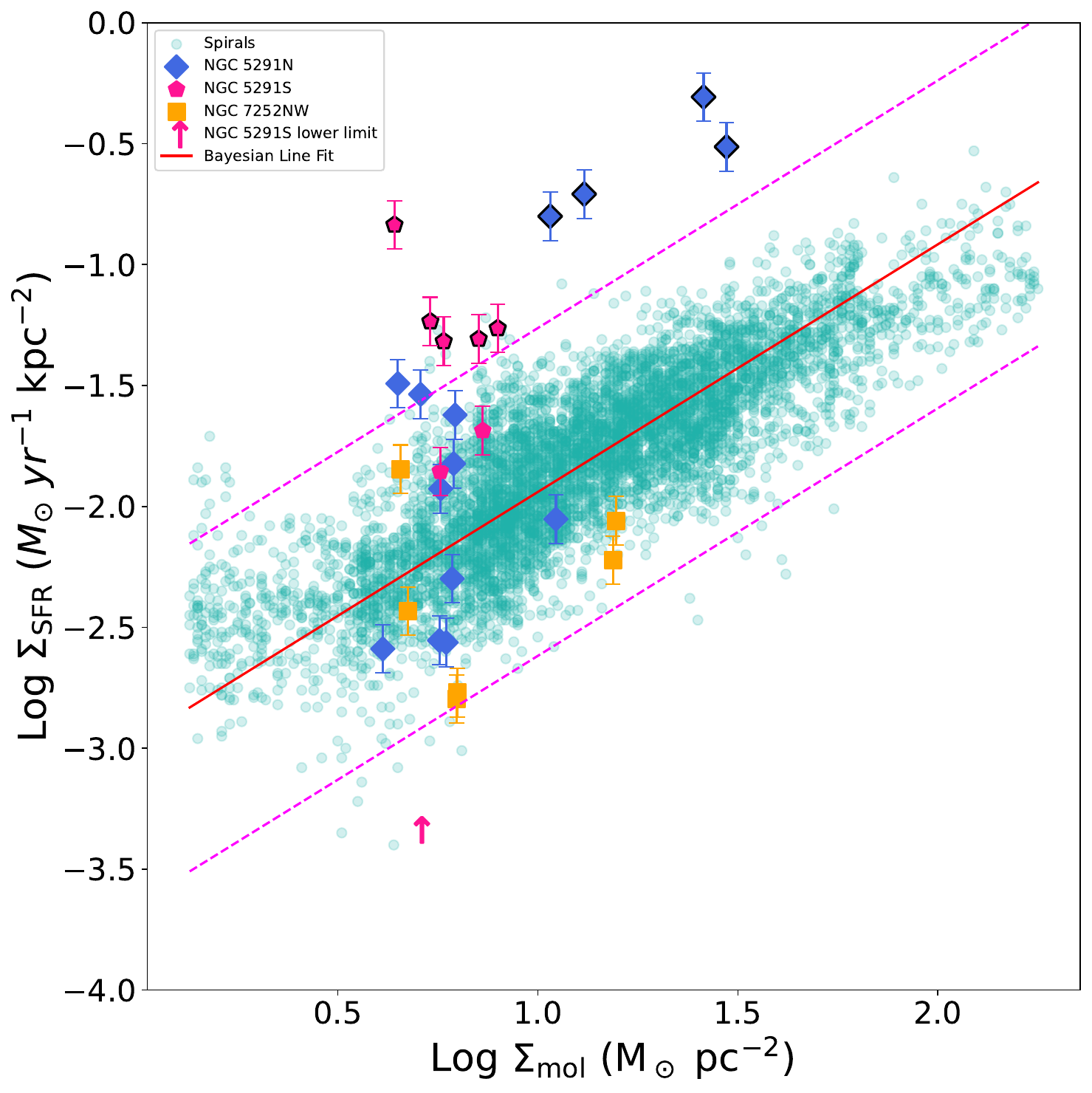}
    \caption{The location of TDGs (blue diamonds, yellow squares, and pink pentagons) on the spatially resolved Kennicutt-Schmidt relation from the ALMaQUEST survey (cyan circles, from \citealt{lin2019almaquest}). The solid red line shows the best-fit line to the ALMaQUEST data; the dashed magenta lines correspond to $\pm$3$\sigma_{\rm obs}$ where $\sigma_{\rm obs}=0.23$ dex is the observed vertical scatter. Symbols with a thick border correspond to regions in which young massive star clusters have been identified (see Fig.\,\ref{fig:datapoints}). }
    \label{fig:ksrelation}
\end{figure}

Unfortunately, we cannot study the spatially resolved SF relation considering the total gas surface densities (molecular plus atomic gas) because the angular resolution of the existing \hi\ data of TDGs is too coarse. In fact, the entire CO emitting area is within one \hi\ beam. The ALMaQUEST data, however, do not consider the \hi\ surface densities as well, so our comparison in Fig.\,\ref{fig:ksrelation} is self-consistent.

Figure \ref{fig:ksrelation} shows the location of TDGs on the spatially resolved SF relation from ALMaQUEST. We fit the ALMaQUEST data with a linear relation using the Markov-Chain Monte-Carlo (MCMC) software BayesLineFit \citep[][]{lelli2019baryonic}. The MCMC fit returns a slope of 1.024 $\pm$ 0.008, an intercept of -2.964 $\pm$ 0.009, and a vertical observed scatter $\sigma_{\rm obs}= 0.23$ dex. The majority of TDG points (16/27) lie on the same SF relation as spiral galaxies within $\pm$3$\sigma_{\rm obs}$. The correlation between $\Sigma_{\rm SFR}$ and $\Sigma_{M_{\rm mol}}$, however, is not evident when considering only TDG data. On the one hand, this occurs because most TDG points cover a small dynamic range in gas surface density (less than 1 dex) and the SF relation has substantial scatter at fixed $\Sigma_{\rm mol}$. On the other hand, TDG data display a larger scatter from the fitted line than the ALMaQUEST data: 0.7 dex considering all TDG points and 0.4 dex considering only those within $\pm$3$\sigma_{\rm obs}$. Indeed, only $\sim$30\%-35\% of TDG points lie within $\pm$1$\sigma_{\rm obs}$ of the best-fit relation rather than the expected 68$\%$ for a Gaussian distribution. 

The high scatter of TDG points may be due to small-number statistics or systematic differences between our work and the ALMaQUEST analysis. If real, instead, it could point to (i) the need of considering the total gas surface density (atomic plus molecular gas) in \hi-dominated galaxies, as in the case of the spatially integrated SF relation in Fig.\,\ref{fig:globalKSlaw}, (ii) high stochasticity in the SF history of TDGs on small spatial scales \citep[e.g.,][]{Boquien2010}, (iii) spatial variations in the $X_{\rm CO}$ factor due to additional effects (beyond gas metallicity) such as gas temperature, gas pressure, and UV background \citep[e.g.,][]{bolatto2013co}, (iv) differences in 3D volume densities due to line-of-sight integration and variable disk thickness \citep[e.g.,][]{Bacchini2019, Bacchini2020}. Given the complex evolutionary status of TDGs, which are possibly out of dynamical equilibrium \citep{lelli2015gas}, it is difficult to distinguish between these possibilities.

Interestingly, a substantial fraction of TDG regions (10/28) strongly deviate from the observed SF relation and lie in the starburst zone above $+$3$\sigma_{\rm obs}$. These starburst regions belong to NGC\,5291N and NGC\,5291S. Studies with the Hubble Space Telescope (HST) shows that these regions are currently forming young star clusters with masses ranging from a few $10^{3}$ M$_\odot$ to a few $10^{5}$ M$_\odot$ and ages from $\sim$1 Myr to $\sim$100 Myr \citep{fensch2019massive}. Thus, it is sensible that these areas have exceptionally high SFEs.

Unlike NGC\,5291N and NGC\,5291S, NGC\,7252NW does not show starburst regions with high SFE, but most of its SF regions (4/6) fall below the average SF relation. Consistently, visual inspection of the available HST images of NGC\,7252NW does not reveal any clear young star cluster. Another two TDGs with spatially resolved CO data (VCC\,2062 from \citealt{lisenfeld2016molecular} and  J1023+1952 from \citealt{querejeta2021alma}) were also found to lie systematically below the average SF relation, albeit \citet{querejeta2021alma} warn that the inclusion or exclusion of diffuse CO emission (not contained in giant molecular clouds) could result into a large difference. The different behaviours shown by different TDGs may be related to the evolutionary status of the parent system and the ``age'' of the specific TDG. According to numerical simulations, the gas ring around NGC\,5291 was formed by a head-on galaxy collision about $\sim$360 Myr ago \citep{Bournaud2007}. On the other hand, NGC\,7252 is a late-stage merger resulting from the interaction of two spiral galaxies about $\sim$700 Myr ago \citep{hibbard1995dynamical, chien2010dynamically}. One may speculate, therefore, that the TDGs around NGC\,5291 are young and experiencing a period of peak SF activity due to efficient \hi-to-H$_2$ conversion, whereas those around NGC\,7252 are slightly older and more quiescent. A larger sample of TDGs around more diverse interacting systems, together with detailed numerical simulations of the system, is needed to study the relation between the interaction stage and SF activity in tidal debris.

Broadly speaking, star-forming galaxies can be classified into ``normal'' and ``starbursts'' using a 3$\sigma$ threshold from the mean KS relation. With such a definition, starbursts represent a SF process that occurs with less than 99.7\% chance for a Gaussian distribution of SFEs. The SF regions in our TDG sample display a continuous range of SFEs, but a very large fraction of them ($\sim$40$\%$) proceed in starburst mode, resulting in molecular gas depletion times as short as $10-100$ Myr. In the remaining $\sim$60$\%$ of TDG regions, the SF activity proceeds in a similar way as in normal spiral galaxies, regardless of the different environmental conditions. In this sense, TDGs are ``hybrid'' systems because they contain some regions behaving as normal galaxies and others as starbursts.

\subsection{Timescales and evolution of TDGs}

A spatially resolved KS relation with a slope of one (as observed) implies that the molecular gas depletion time ($t_{\rm mol}=M_{\rm mol}$/SFR) is nearly constant across spiral galaxies. Then, the intercept and observed scatter imply that $t_{\rm mol}\simeq1\pm0.5$ Gyr. For our three TDGs, the molecular gas depletion times are substantially smaller: 100 Myr for NGC\,5291N, 70 Myr for NGC\,5291S, and 300 Myr for NGC\,7252NW. The molecular gas of these TDGs, therefore, will soon be consumed by the SF activity unless it is replenished by efficiently converting the substantial \hi\ reservoir into H$_2$ gas.

Considering the total \hi\ mass associated with the TDG potential well \citep[][their Table 8]{lelli2015gas}, the atomic gas depletion time ($t_{\rm atom}=M_{\rm atom}$/SFR) is about 2 Gyr for NGC\,5291N, 4 Gyr for NGC\,5291S, and 7 Gyr for NGC\,7252NW. These values of $t_{\rm atom}$ are substantially smaller than those of low-surface-brightness (LSB) star-forming galaxies, ranging between $10-100$ Gyr \citep[e.g.,][]{mcgaugh2017}, but are comparable to those of starburst dwarf galaxies \citep{lelli2014evo}, such as blue compact dwarfs (BCDs). It is conceivable that both TDGs and BCDs are only able to sustain the intense SF activity for a short period of time ($\sim$0.5-1 Gyr, e.g., \citealt{mcquinn2010}), so the starburst will not have enough time to consume their entire \hi\ reservoir. For example, if their SFR is going to decrease in the next 500 Myr by a factor of $\sim$10, their $t_{\rm atom}$ will increase by a similar factor, reaching the high values observed in LSB galaxies. Furthermore, there is a diffuse \hi\ reservoir in the tidal debris around the TDGs, which might replenish their gas content.

Another interesting timescale is the stellar mass growth time ($t_\star = M_\star$/SFR), which we compute using the stellar masses from \citet[][their Table 8]{lelli2015gas}. For NGC\,7252, we find $t_\star\simeq940$ Myr. This is larger than the dynamical timescale of the galaxy merger ($\sim$700 Myr) inferred from numerical simulations \citep{chien2010dynamically}. Assuming that the age of the TDG is equal to that of the merger, the current SFR cannot explain the present-day stellar mass: the SFR was probably higher in the past, indicating that the SF activity has been declining over time. A possible caveat is that NGC\,7252NW may contain old stars from the disk of the parent galaxies, which would contribute to $M_\star$ beyond the mass formed over the past 700 Myr. In any case, the situation of NGC\,5291 appears different: we find $t_\star \simeq 180$ Myr for NGC\,5291N Myr and $t_\star\simeq270$ Myr for NGC\,5291S, which are smaller than the dynamical timescale of the galaxy collision ($\sim$360 Myr, \citealt{Bournaud2007}). Thus, the current SFR can amply explain the present-day $M_\star$ of these two TDGs. These facts are in line with the speculation in Sect.\,\ref{sec:ksrelation} that the TDGs around NGC\,5291 may be representative of early galaxy formation with efficient \hi-to-H$_2$ conversion, high SFE, and short gas depletion times, while those around NGC\,7252 may represent a subsequent stage with more typical SF activity. In addition, the initial conditions in the two systems may have been different: while the TDGs around NGC5291 were born out of pure gaseous condensation, those around NGC7252 may have been born in a less \hi\ dominated environment with both gas and stars from their progenitors.

\section{Conclusions}\label{sec:conc}

We studied the molecular and ionized gas content of three bona-fide TDGs using CO($1-0$) observations from ALMA and IFS data from MUSE. For the first time, we locate TDGs on the spatially resolved KS relation. Our results can be summarized as follows:

\begin{enumerate}
    \item CO($1-0$) and H$\alpha$ emissions in TDGs are very compact and cover a much smaller area than the \hi\ emission. Both CO and H$\alpha$ lines are not suitable to study the internal kinematics of these TDGs. Most likely, molecular gas is forming out of the more extended \hi\ disk of TDGs, having similar line-of-sight velocities.
    \item TDGs lie on the same spatially-integrated $\Sigma_{\rm SFR}-\Sigma_{\rm gas}$ relation of spiral galaxies but display a substantial scatter on the spatially resolved $\Sigma_{\rm SFR}-\Sigma_{\rm mol}$ relation (which neglects atomic gas due to the lack of high-resolution \hi\ data).
    \item The majority (60$\%$) of SF regions in TDGs lie on the same spatially resolved SF relation as spiral galaxies within $\pm$3 times the observed scatter but display a larger dispersion from the mean relation. A substantial fraction ($\sim$40$\%$) of SF regions have exceptionally high SF efficiencies, lying in the starburst regime of the KS relation. These regions belong to NGC\,5291N and NGC\,5291S, and are associated with the formation of massive super star clusters, which were previously identified by HST imaging \citep{fensch2019massive}.
\end{enumerate}

The growing evidence about the existence of a fundamental SF relation compels us to put the relation to test in a variety of star-forming environments. The three TDGs analyzed here confirm the fundamental nature of the KS relation on sub-kpc scales, albeit regions with exceptionally high SF efficiencies do exist. Future studies may investigate the spatially resolved KS relation in other bona-fide TDGs, probing even more diverse star-forming environments, such as younger and/or older tidal debris.

\section*{Acknowledgements}
We are grateful to the anonymous referee for the useful comments that helped to improve the paper. We thank Rob Kennicutt and Lin Li-Hwai Lin for providing the KS-relation data in tabular form. N.K. and F.L. thank the School of Physics and Astronomy of Cardiff University, where this work started as part of a Master thesis project. N. K. acknowledges support from the Programa de doctorado en Astrofísica y Astroinformática of Universidad de Antofagasta. M.B. acknowledges support from FONDECYT regular grant 1211000 and by the ANID BASAL project FB210003. U.L. acknowledges support by the research projects AYA2017-84897-P and PID2020-114414GB-I00 from the Spanish Ministerio de Econom\'\i a y Competitividad, from the European Regional Development Funds (FEDER) and the Junta de Andaluc\'ia (Spain) grants FQM108. This paper makes use of the following ALMA data: ADS/JAO.ALMA\#2015.1.00645.S. ALMA is a partnership of ESO (representing its member states), NSF (USA) and NINS (Japan), together with NRC (Canada), MOST and ASIAA (Taiwan), and KASI (Republic of Korea), in cooperation with the Republic of Chile. The Joint ALMA Observatory is operated by ESO, AUI/NRAO and NAOJ. Based on observations collected at the European Organisation for Astronomical Research in the Southern Hemisphere under ESO MUSE program 60.A-9320(A) and 097.B-0152(A). Based on public data released from the MUSE WFM-AO commissioning observations at the VLT under Programme IDs 60.A-9100 runs G \& H.

\section*{Data Availiability}

The raw data used in this work are available in the ALMA and ESO archives. Reduced data are available on request.




\bibliographystyle{mnras}
\bibliography{bibilography} 








\label{lastpage}
\end{document}